\documentclass[twocolumn,showpacs,preprintnumbers,amsmath,amssymb]{revtex4}

\usepackage{epsfig}
\usepackage{dcolumn}
\usepackage{bm}
\usepackage{latexsym}

\begin{document}

\title{Stimulated emission from the biexciton in a single quantum dot }

\author{I.~A.~Akimov, J.~T.~Andrews, and F.~Henneberger}
\affiliation{ Humboldt Universit\"at zu Berlin, Institut f\"ur
Physik, Newtonstr.15, 12489 Berlin, Germany}

\date{\today}

\begin{abstract}
Using two optical pulses of different frequencies, we demonstrate
entanglement and disentanglement of the electronic states in
Stranski-Krastanov quantum dots. Resonant two-photon excitation of
the biexciton creates an entangled Bell-like state. The second
pulse, being resonant to the exciton-biexciton transition,
stimulates the emission from the biexciton and fully disentangles
the two-bit system. By setting the polarization of the stimulation
pulse, we control the recombination path of the biexciton and, by
this, the state of the photons emitted in the decay cascade.
\end{abstract}



\maketitle

Entangled states and their manipulation play an important role in
quantum information processing \cite{book}. A state of a pair of
quantum systems is said to be entangled if it cannot be factored
into the states of the individual subsystems. When aiming at
practical devices, implementations in solid state are of
particular interest. Here, a critical issue is decoherence that
destroys the non-classical quantum correlations.

\begin{figure}
 \begin{minipage}{8.2cm}
  \epsfxsize=8 cm
  \centerline{\epsffile{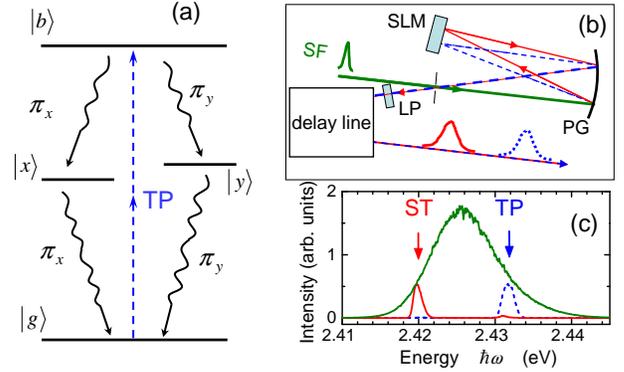}}
  \caption{\label{fig:Fig1}
(a) Schematics of the exciton-biexciton system in a QD. For
explanations see text. (b) Layout of the pulse shaper. SF: sub-ps
source pulse, SLM: spatial light modulator, PG: parabolic grating,
LP: linear polarizer. (c) Spectra of the ingoing SF pulse and the
resulting ST and TP pulse.}
  \end{minipage}
\end{figure}

Semiconductor quantum dots (QDs) with electronic excitations
localized on a nanometer length scale have recently attracted much
attention as building blocks for quantum logic \cite{Steel,Li}.
The two-exciton subspace of a QD is spanned by the ground-state
$|\text{g}\rangle$ (no exciton), the single-exciton states, and
the biexciton state $|\text{b}\rangle$. The optically active
exciton is split by anisotropy in two linearly cross-polarized
components $|\text{x}\rangle$ and $|\text{y}\rangle$
\cite{Gammon-Bacher}. Thus, identifying
$|00\rangle=|\text{g}\rangle$, $|10\rangle= |\text{x}\rangle$,
$|01\rangle=|\text{y}\rangle$, and $|11\rangle=|\text{b}\rangle$,
a two-bit system is built. The optical couplings between those
states form the $\text{V}-\Lambda$ transition scheme of
Fig.~\ref{fig:Fig1}a, where the exciton-biexciton resonances are
low-energy shifted from that of the single excitons by the
exciton-exciton interaction energy $\Delta E_\text{XX}$ in the
biexciton.

Coherent optical coupling of the ground-state and the biexciton
creates an entangled state of the type
$a_{00}|\text{00}\rangle+a_{11}|\text{11}\rangle$
\cite{Hohenester,Biolatti-Sham}. A Bell state of maximum
entanglement is achieved if $|a_{ii}|= \frac{1}{\sqrt{2}}$.
Excitation of the biexciton
$|\text{g}\rangle\rightarrow|\text{b}\rangle$ requires two
photons. In Ref.\cite{Chen}, two non-degenerate beams, each being
resonant to one of the single-photon transitions, e.g.
$|\text{g}\rangle\rightarrow|\text{x}\rangle$ and
$|\text{x}\rangle\rightarrow|\text{b}\rangle$, have been utilized.
The present study is based on resonant two-photon (TP) excitation
\cite{TPCC}. That approach, where the degenerate photons have half
the energy of the biexciton, has the advantage that the
spontaneous emission from the cascaded biexciton-exciton decay is
clearly separated from the excitation stray light and can thus be
used as a monitor of the quantum dynamics. The ability to track
the QD emission is essential, as it enables to generate
non-classical light states, like anti-bunched single-photon
emission \cite{Michler-Santori} or entangled photon pairs
\cite{Benson}.

In what follows we demonstrate that an entangled state in the
two-exciton subspace of a QD can be indeed created by resonant TP
excitation. The entanglement of formation reaches values of about
0.5. In a next step, we accomplish the conversion of the biexciton
into an exciton by stimulated emission applying a second pulse
that is tuned to the exciton-biexciton resonance. In this way, the
recombination path is strictly defined by the polarization of the
stimulation pulse, in contrast to the spontaneous emission
cascade. In a quantum information sense, the stimulated emission
corresponds to a disentanglement of the Bell-like state
$a_{00}|\text{00}\rangle+a_{11}|\text{11}\rangle \rightarrow
(a_{00}|\text{0}\rangle+a_{11}|\text{1}\rangle)|0\rangle$.
Disentanglement is a key step in conditional quantum dynamics and
logical gates \cite{Barenco}. We find efficiencies of close to 1
in our measurements.

The Stranski-Krastanov CdSe/ZnSe QD structures are grown by
molecular beam epitaxy \cite{Litvinov}. II-VI QDs are favored in
the present context by the large $\Delta E_\text{XX} \approx$
20~meV \cite{Kulakovski-Kreller}, allowing the use of ultra-short
pulses without loosing spectral selectivity of the excitation
process. In order to study individual QDs, mesa structures with an
area down to $100\times100$~nm$^2$ are fabricated. The sum
frequency of a Kerr-lens mode-locked Ti:sapphire laser and a
synchronously pumped optical parametric oscillator are used to
generate spectrally broad sub-ps pulses with 76~MHz repetition
rate in the spectral region of interest. Two spectrally narrow
pulses for selective TP excitation and stimulation (ST) of the
biexciton are obtained by a pulse shaper based on a programmable
reflective spatial light modulator (Fig.~\ref{fig:Fig1}b). The
outgoing pulses are spatially separated and passed through a delay
line. The pulse durations are about 1.0~ps (TP) and 1.5~ps (ST)
and the spectral full widths at half maximum are 1.7~meV and
1.2~meV, respectively (Fig.~\ref{fig:Fig1}c). The secondary
emission of the QD is collected in a confocal arrangement and
dispersed in a triple spectrometer (0.23~nm/mm) equipped with a
nitrogen cooled charged coupled device. For time-resolved
measurements, only the two first stages are utilized in
subtractive mode (0.7~nm/mm). A multi-channel-plate
photomultiplier in conjunction with time-correlated single-photon
counting unit provides an overall time resolution of 60~ps.
Polarization control is achieved by quarter-wave or half-wave
plates, placed in the path of both the linearly polarized
excitation light as well as the emission signal. The polarization
of the emission is analyzed by a Glan-Thomson prism introduced in
front of the spectrometer. All measurements are carried out at
temperatures of about 10~K.

\begin{figure}
 \begin{minipage}{8.2cm}
  \epsfxsize=8 cm
  \centerline{\epsffile{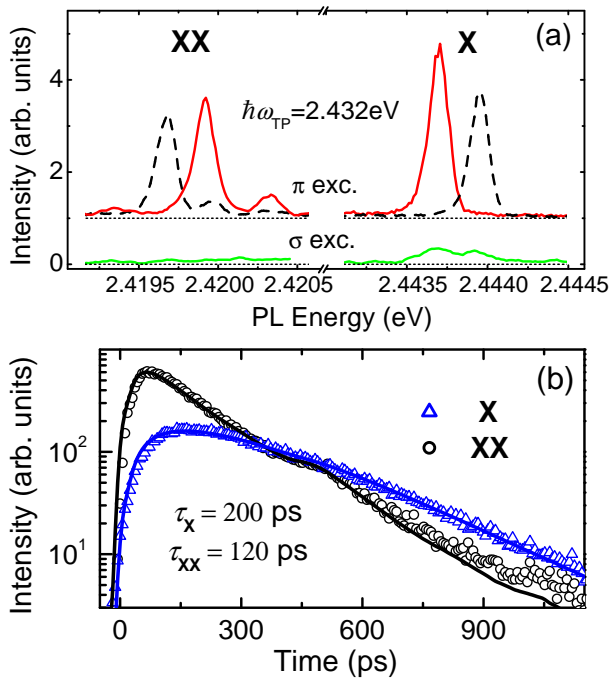}}
  \caption{\label{fig:Fig2}
(a) Exciton (X) and biexciton (XX) emission lines from a single QD
under circularly ($\sigma$) and linearly ($\pi$) polarized TP
excitation. The polarization detection is along (solid) and
perpendicular (dashed) to the intrinsic polarization
$\vec{\pi}_\text{x}$. (b) Decay transients of biexciton and
exciton emission under TP excitation. Solid lines are a
double-exponential data fits accounting for the apparatus
function.}
 \end{minipage}
\end{figure}

The cascaded spontaneous emission of a QD subsequent to TP
excitation is depicted in Fig.~\ref{fig:Fig2}. For linearly
polarized excitation, distinct exciton and biexciton features,
placed symmetrically to the excitation photon energy, are present.
In full accord with the transition scheme of Fig.~\ref{fig:Fig1}
a, both features consist of a fine structure doublet of linearly
cross-polarized lines, however, with a reversed sequence of the
polarization in the exciton and biexciton emission. The TP
transition of the biexciton is forbidden for circular excitation
polarization. While the emission yield decreases indeed by one
order of magnitude, a weak rest emission at the exciton survives.
This background originates from electron-hole pairs off-resonantly
excited in the energy continuum of the hetero-structure and
captured by the QD. On the other hand, under linearly polarized
excitation, the emission signal at the exciton lines is entirely
insensitive on the polarization direction, excluding that the TP
pulse addresses directly the exciton states to a measurable
extent. A flip between $|\text{x}\rangle$ and $|\text{y}\rangle$
takes place on a time-scale markedly longer than the life-time and
can be thus ignored \cite{Flissi}. The time-resolved emission
shown in Fig.~\ref{fig:Fig2}b clearly confirms the existence of an
emission cascade, with the biexciton recombining first and with a
respective rise time for the exciton. Double-exponential fits
yield that the radiative life-time of the biexciton
($\tau_\text{XX}$ = 120 ps) is about two times shorter than for
the exciton ($\tau_\text{X}$ = 200 ps). Since the biexciton has
two recombination channels this means that the exciton and the
exciton-biexciton transition posses almost the same dipole moment.
Using $d^2=3 \pi\varepsilon_0\hbar c^3/n_\text{r} \omega_0^3
\tau$, we find $d=$ 27 Debye.

\begin{figure}
 \begin{minipage}{8.2cm}
  \epsfxsize=7.5cm
  \centerline{\epsffile{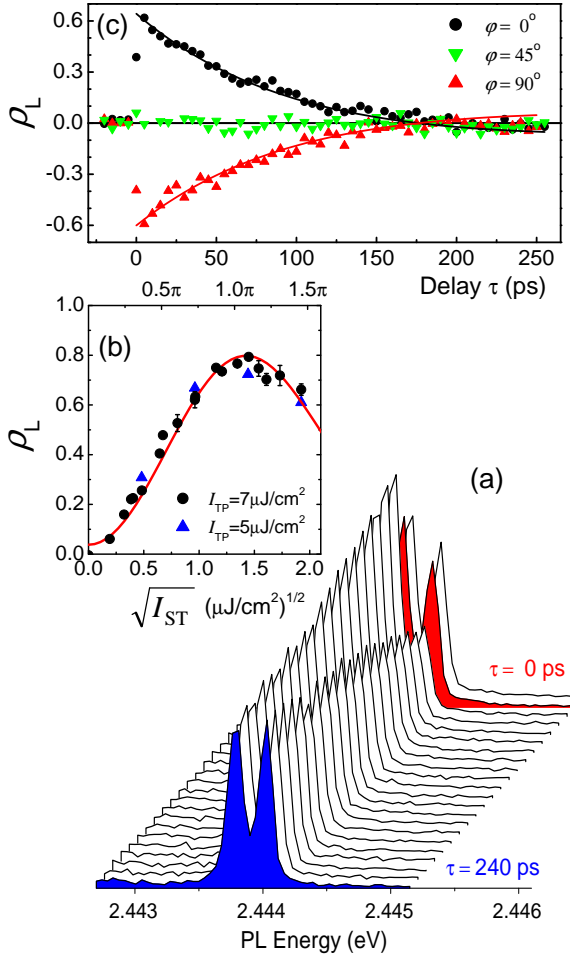}}
  \caption{\label{fig:Fig3}
(a) Evolution of exciton emission spectrum as a function of the
delay $\tau$ between TP and ST pulse. The ST polarization is along
the polarization of the low-energy line. (b) Induced linear
polarization degree $\rho_\text{L}$ versus square-root
pulse-density $\sqrt{I_\text{ST}}$ ($\tau = 5~\text{ps}, \varphi =
0$). (c) $\rho_\text{L}$ versus $\tau$ for ($I_\text{TP} = 7~\mu
\text{J/cm}^2, I_\text{ST} = 1.3~\mu \text{J/cm}^2$),
$\varphi=\angle (\vec{e},\vec{\pi}_\text{x})$. Since not related
to the intrinsic QD dynamics, the off-resonant background is
subtracted in (b) and (c). The solid lines are fits to the data
based on the numerical solution of the Master equation for the
two-exciton density matrix. For details see text.}
  \end{minipage}
\end{figure}

\begin{figure}
 \begin{minipage}{8.2cm}
  \epsfxsize=7.5cm
  \centerline{\epsffile{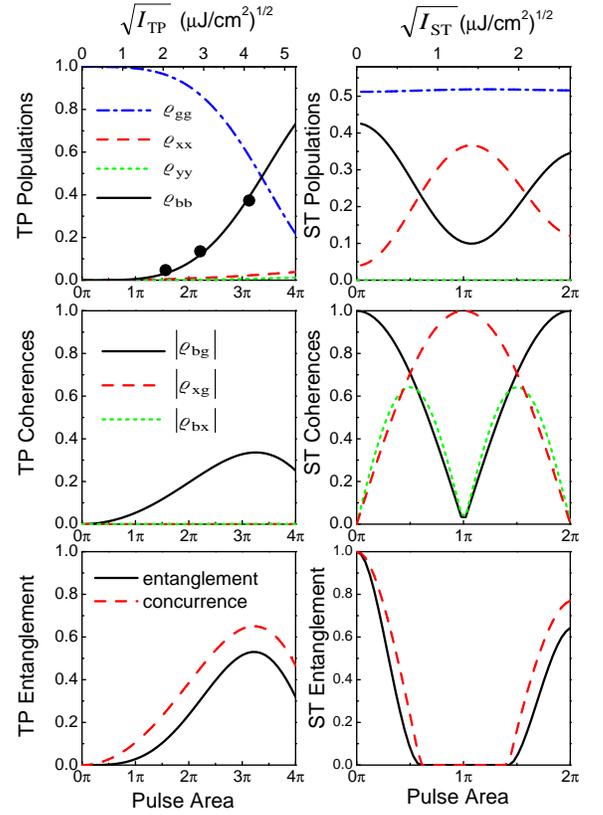}}
  \caption{\label{fig:Fig4}
Calculated populations, coherences, entanglement, and concurrence
versus pulse area for the TP (left) and the ST pulse (right). The
experimental points in the left top panel represent the normalized
signal of the XX emission under TP excitation only. The plots in
the two lower-right panels are normalized with respect to the
initial values produced by the TP pulse. $\tau$ = 5~ps,
$\theta_\text{TP}$ = 3 $\pi$.}
  \end{minipage}
\end{figure}

Now we focus on the stimulated emission of the biexciton. In these
measurements, the QD is excited by a sequence of TP and ST pulses,
linearly co-polarized under an angle $\varphi$ relative to the
intrinsic polarization $\vec{\pi}_\text{x}$ and with tuneable
time-delay $\tau$. The result of the stimulation process is a
photon as well as an exciton. While the photon can hardly be
detected, the stimulated exciton is manifested by the photon that
it emits subsequently. The data in Fig.~\ref{fig:Fig3}a directly
verify this scenario. While both exciton emission components have
equal intensity for TP excitation only, the line polarized along
the ST polarization is amplified at expense of the cross-polarized
line when both pulses are present. During pulse delay, the
biexciton state is increasingly emptied by spontaneous decay.
Consistent with the radiative life-time, the transition is not
longer capable of stimulated emission after about 250 ps. A
measure to what extend a certain exciton state can be selected by
the ST pulse is given by the induced linear polarization degree
$\rho_\text{L}=(I_\text{x}-I_\text{y})/(I_\text{x}+I_\text{y})$,
$I_{i}$ being the spectrally integrated signal of $|i\rangle$. As
seen in Fig.~\ref{fig:Fig3}c, $\rho_\text{L}=0$ for
$\varphi=\pi/4$, while, exciting along the intrinsic polarization
axis, $\rho_\text{L}$ has the same absolute value, but is positive
for $\varphi=0$ and negative for $\varphi=\pi/2$. In an incoherent
regime, the polarization degree is limited to $\rho_\text{L}=0.5$,
because the inversion between the biexciton and exciton
population, addressed for a given ST polarization, can not exceed
zero. In contrast, $\rho_\text{L}$ clearly exceeds this limit and
reaches values of up to 0.8 for the maximum pulse densities
available by our set-up. Even the onset of Rabi oscillations for
the exciton-biexciton transition is evidenced in
Fig.~\ref{fig:Fig3}b. The relatively large fine structure
splitting of the exciton allows us to spectrally separate
spontaneous and stimulated emission. For QDs where the splitting
is within the homogeneous width the stimulation process can be
also accomplished, but shows only up in the polarization degree.

In order to draw quantitative conclusions on the quantum dynamics
behind the experimental observations, the full density matrix of
the two-exciton subspace has to be considered. Its time evolution
is determined by the Master equation $\imath\hbar\dot{\varrho}=[
H,\varrho ]+\imath\hbar\Gamma[ \rho ]$ with the Hamiltonian given
in units of $\hbar$ and rotating-wave approximation by
\begin{eqnarray}
\label{eq:Hamiltonian} H &=& \omega_\text{g}|\text{g} \rangle
\langle \text{g}| +\omega_\text{x}|\text{x} \rangle\langle
\text{x}|+\omega_\text{y}|\text{y} \rangle\langle \text{y}|)+
\omega_\text{b}|\text{b} \rangle\langle \text{b}| \nonumber\\
 & &-\frac{1}{2}\{ \Omega(t)\text{e}^{-\imath \omega t}[ \cos(\varphi)(|\text{x}\rangle\langle
\text{g}|+|\text{b} \rangle \langle \text{x}|)\nonumber\\
& &+ \sin(\varphi)(|\text{y}\rangle
\langle\text{g}|+|\text{b}\rangle \langle
 \text{y}|)]+\text{h.c.}\nonumber
\end{eqnarray}
$\Omega=\frac{d}{\hbar} \mathcal{E}(t)$ is the time-dependent Rabi
frequency of the field amplitude $\mathcal{E}(t)$, whereby the
same $d$ is taken for all transitions. Expressed in the
two-exciton basis, the Master equation defines a complete set of
differential equations for the 10 independent density matrix
elements $\varrho_{ij}=\varrho_{ji}^\ast$. For the relaxation
terms, we use $\langle i|\Gamma[\varrho]|j\rangle =
-\Gamma_{ij}\varrho_{ij}~(i \neq j),
\langle\text{b}|\Gamma[\varrho]|\text{b}\rangle =
-\varrho_\text{bb} /\tau_\text{XX}, \langle i| \Gamma[\varrho]|
i\rangle = \varrho_\text{bb}/ 2 \tau_\text{XX}
-\varrho_{ii}/\tau_\text{X}~(i = \text{x,y})$, and
$\dot{\varrho}_\text{bb}+\dot{\varrho}_\text{xx}+\dot{\varrho}_\text{yy}+\dot{\varrho}_\text{gg}=0$.
Except the off-diagonal damping rates, all parameters $(\Delta
E_{\text{xx}}, E_\text{x}-E_\text{y}, d, \tau_\text{XX},
\tau_\text{X})$ entering the equations are known experimentally.
Therefore, though the set is relatively large, the population
dynamics is fully defined and the $\Gamma_{ij}$, describing the
coherence decay, can be deduced by comparing with the experimental
data. The dynamical response is independent on whether
$|\text{x}\rangle$ or $|\text{y}\rangle$ is addressed. Hence,
$\Gamma_\text{xg}=\Gamma_\text{yg}$,
$\Gamma_\text{bx}=\Gamma_\text{by}$, and making the reasonable
assumption $\Gamma_\text{bx}=\Gamma_\text{xg}+\Gamma_\text{bg}$,
the only two free parameters left are the exciton
($\Gamma_\text{xg}$) and biexciton ($\Gamma_\text{bg}$)
decoherence rates.

We have numerically solved the equations for Gaussian pulse shapes
assuming $\varrho_\text{gg}=1$ and all other $\varrho_{ij}=0$
before the TP pulse and taking the resultant $\varrho_{ij}$ as
initial values for the interaction with the ST pulse. In a
time-integrated detection mode, the linear polarization degree is
given by $\rho_\text{L}=(\delta \varrho_\text{xx}-\delta
\varrho_\text{yy})/(\delta \varrho_\text{bb}+\delta
\varrho_\text{xx}+\delta \varrho_\text{yy})$, $\delta
\varrho_{ii}$ denoting the change of the population generated by
the pulses. Calculating $\rho_\text{L}$ as a function of the pulse
delay yields indeed perfect agreement with experimental curves in
Fig.~\ref{fig:Fig3}c, verifying that the population dynamics is
governed by the life-times $\tau_\text{XX}$ and $\tau_\text{X}$.
The decoherence rates follow from the density dependence of
$\rho_\text{L}$ (Fig.~\ref{fig:Fig3}b). Too large rates spoil
rapidly the purity of the TP excitation, resulting in a
significant polarization degree even without the ST pulse, while
the maximum level of only $\rho_\text{L}=0.8$ signifies the
presence of pure dephasing beyond the radiative damping. The fit
to the data yields $\Gamma^{-1}_\text{xg}\approx
\Gamma^{-1}_\text{bg} = 6$ ps. A careful analysis of the spectral
line-shape of the exciton emission provides that the radiative
Lorentzian is superimposed to a weak but broad acoustic phonon
background, which translates in non-exponential damping in the
time domain with a short component consistent with the rates
deduced from the ST data (see also \cite{Borri-Besombes}).
Distinct non-exponential damping is also indicated by TP coherent
control measurements \cite{TPCC}. Here, a significant drop of the
contrast occurs right after pulse separation, whereas the
subsequent decay evolves on a much longer time-scale.
Fig.~\ref{fig:Fig4} represents plots of the density matrix
elements versus pulse area $\theta = \int_{-\infty}^{+\infty}
\Omega(t)dt$ in the experimentally relevant pulse-density range.
Despite of the relatively fast dephasing, the TP pulse creates a
significant biexciton coherence $\varrho_\text{bg}$ that can be
manipulated with the succeeding ST pulse. Damped Rabi oscillations
have been already observed previously \cite{Steel,Li,Rabi}.
However, unlike our study, where we monitor an absolute quantity,
no conclusions about the true coherence degree could be made.

Knowing the whole density matrix, the entanglement of formation
$E(\varrho)$ can be calculated. Following Ref.\cite{Wootters},
$E=E(C)=h(\frac{1+\sqrt{1-C^2}}{2})$ with the concurrence $C$ and
$h(x)=-x \log_2 (x)-(1-x)\log_2 (1-x)$. For a purely coherent
regime, it is straightforward to show that $C=2
\sqrt{\varrho_\text{gg}\varrho_\text{bb}}$. While we can come with
the TP pulse experimentally close to a situation
$\varrho_\text{bb} \approx \varrho_\text{gg} \approx \frac
{1}{2}$, the actual entanglement is markedly lower than 1.
Decoherence establishes a mixed state, where now
$C(\varrho)=\text{max}[0,\lambda_1-\lambda_2-\lambda_3-\lambda_4]$,
the $\lambda$'s being in decreasing order the square roots of
$\varrho
\sigma_y\otimes\sigma_y\varrho^\ast\sigma_y\otimes\sigma_y$ with
the time inversion operator $\sigma_y$ \cite{Wootters}. The
calculation provides a maximum entanglement of about 0.5. However,
this limited value given, complete disentanglement of the
2-quantum-bit system is accomplished by the ST pulse in a wide
range of pulse areas (Fig.~\ref{fig:Fig4}).

In conclusion, we have demonstrated entanglement and
disentanglement of the electronic states of a QD by a sequence of
two optical pulses. Being able to control the recombination path
of the biexciton, the cascade emission can be adjusted from a pair
of correlated or even entangled photons to a single photon of
defined polarization or in an entangled polarization state. While
an extrinsic background spoiling the figure-of-merit can be
eliminated by appropriate sample design, our measurements uncover
also rather large decoherence rates, probably related to a
non-Markovian acoustic phonon contribution. This point deserves
further investigations beyond the scope of this work. Our results
are also of relevance for the use of QD in lasers, as they show
that the exciton level is rapidly emptied avoiding reabsorption.

The authors thank S. Rogaschewski for the lithographic etching.
This work was supported by the Deutsche Forschungsgemeinschaft
within Project No. He 1939/18-1.

\end{document}